\documentclass[11pt, oneside]{article}   	
\usepackage{geometry}                		
\usepackage{graphicx}				
\usepackage{amssymb}
\usepackage{amsmath}
\usepackage{subfig}
\usepackage{caption}
\usepackage{comment}

\usepackage[utf8]{inputenc}
\usepackage{physics}
\usepackage{braket}
\usepackage{amsmath}
\numberwithin{equation}{section}
\begin{document}
\title{Morse Potential on a Quantum Computer for\\ Molecules and  
Supersymmetric Quantum Mechanics}
\author{Josh Apanavicius, Yuan Feng, Yasmin Flores, Mohammad Hassan, Michael McGuigan\footnote{In author order: Indiana University, Pasadena City College, St. Joseph's College, City College of NY,  Brookhaven National Laboratory}}
\date{}
\maketitle
\begin{abstract}
In this paper we discuss the Morse potential on a quantum computer. The Morse potential is useful to describe diatomic molecules and has a finite number of bound states which can be measured through spectroscopy. It is also a example of an exactly soluble potential using supersymmetric quantum mechanics. Using the the supersymmetric quantum mechanics formalism one can derive a heirachy of Hamiltonians such that the ground state of the next rung on the heirarchy yeids the first excited state of the hamiltonian below it. Using this method one can determine all the states of the Morse potential by calculating all the ground states of the sequence of Hamiltonians in the heirarchy. We use the IBM QISKit software together with the Variational Quantum Eiegensolver (VQE) algorithm to calculate the ground state and first excited state energy of the Morse potential and find agreement with the exact expression for the bound state energies of the Morse Potential. We analyze different optimizers to study the numerical effect on the calculations. Finally we perform  quantum computations for  diatomic and triatomic molecules to illustrate the application of these techniques on near term quantum computers and find excellent agreement with experimental data. 
\end{abstract}

\newpage

\section{Introduction}

The Morse potential was introduced in \cite{Morse:1929zz} to describe the bound state energies of diatomic molecules. The potential also serves as example like the simple harmonic oscillator as an exactly soluble model in that one can analytically determine all the eigenstates and eigenvalues. The exact solvablity can be traced back to it's relation to supersymmetric quantum mechanics (SusyQM)\cite{Witten:1981nf}\cite{Cooper:1994eh}\cite{Gangopadhyaya:2011wka}\cite{Cooper}\cite{Junker}\cite{Arik}\cite{Berrondo} , definition of ladder operators and a set of Hamiltonians that are related by change of parameters in the potential. The Morse potential takes a different form depending on whether one is studying diatomic molecules or one is interested in the supersymmetric quantum mechanics connection. We will start with the SusyQM form but will return to the diatomic molecule form in a later section. The use of the Variational Quantum Eigensolver (VQE) has been shown to be an efficient quantum algorithm for the calculation of ground state energies on noisy intermediate scale quantum computers. Having an exact solution for the Morse potential gives us an excellent point of comparison and allows us to see what level of accuracy can be achieved on current quantum computing hardware and software.

This paper is organized as follows. In section 1 we give an introduction to the study of the Morse potential on a quantum computer. In section 2 we give the exact solutions to the Morse potential and we go over the relation of the Morse potential to supersymmetric quantum mechanics, discuss the heirarchy of Hamiltonians for the Morse potential and how this can be used to calculate all the bound states,and give the exact solutions to the Morse potential. In section 3 we discuss the calculation of the bound states for the Morse potential using the IBM QISKit eigensolver applied to SUSY partner Hamiltonians,  compute ground state and first excited state energies and compare our results to the exact calculation. In section 4 we discuss the application of the variational quantum eigensolver to calculating the ground state energies of realistic diatomic molecules and in section 5 we discuss the calculations with Morse potentials of two variables that can be used to describe triatomic molecules. Finally in section 6 we state the main conclusions of the paper.

\section{Morse potential}

In it's relation to supersymmetric quantum mechanics the Morse potential is written
as:
\begin{equation} V_{-}(x,A) = e^{-2x}-(2A+1)e^{-x}+A^2\end{equation}
with $A$ given by the Morse parameter. This potential is shown in Figure 1 for $A=5$.
\begin{figure}
\centering
  \includegraphics[width = .5 \linewidth]{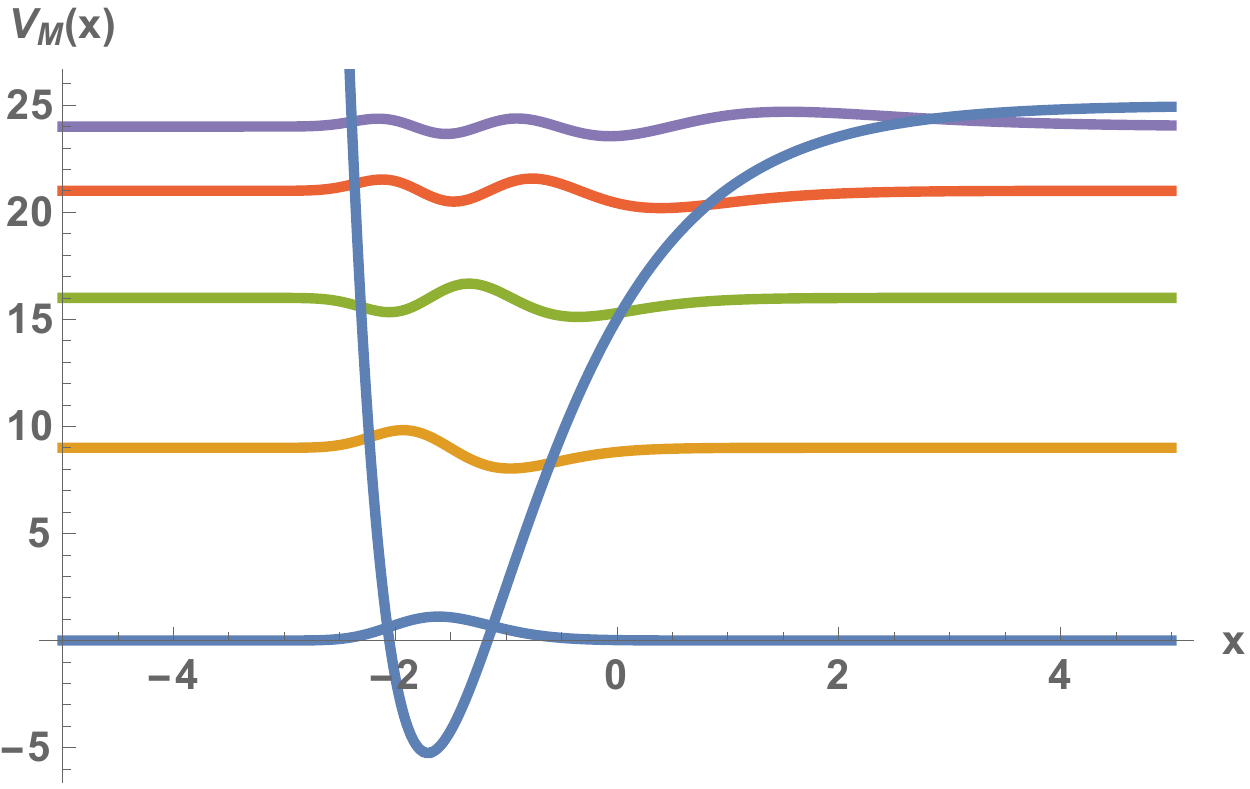}
  \caption{Morse potential for $A=5$ with five bound state energies ${0,9,16,21,24}$}
  \label{fig:Radion Potential}
\end{figure}
When one studies diatomic molecules one writes the Morse potential in a different form:
\begin{equation} V(x) = D(e^{-2a(x-x_0)}- 2e^{-a(x-x_0)})\end{equation}
with parameters $D$, $a$ and $x_0$.

To make the connection with supersymmetric quantum mechanics we write the minus and plus partner potential as:
$${V_ - }(x) = W{(x)^2} - W'(x)$$
 \begin{equation}{V_ + }(x) = W{(x)^2} + W'(x)\end{equation}
 where the superpotential $W(x)$ is given by:
 \begin{equation}W(x) = A - e^{-x}\end{equation}
 so that the plus partner potential is given by:
 \begin{equation} V_{+}(x,A) = e^{-2x}-(2A-1)e^{-x}+A^2\end{equation}
The partner Hamiltonians are:
$$H_{-} = p^2 + {V_ - }(x) = p^2+W{(x)^2} - W'(x)$$
 \begin{equation}H_{+} = p^2 +{V_ + }(x) = p^2 + W{(x)^2} + W'(x)\end{equation}
 where we have set $2m=1$. The ladder operators are given by:
 $$a = i p + W(x) =  i p + A-e^{-x}$$
 \begin{equation}a^{\dagger} = -i p + W(x) =-i p + A -e^{-x} \end{equation}
Then the partner Hamiltonians can be realized as:
$$H_{-} =a^{\dagger} a$$
 \begin{equation}H_{+} =a a^{\dagger}\end{equation}
One can then form a sequence of hierarchy of Hamiltonians as $H_{i}= p^2 + V_i(x)$ where:
\begin{align}
&V_0(x)=V_{-}(x,A)\nonumber\\
&V_1(x)=V_{+}(x,A)\nonumber\\
&V_2(x)=V_{+}(x,A-1)+2(A-1)+1\nonumber\\
&V_3(x)=V_{+}(x,A-2)+2(A-1)+1+2(A-2)+1\nonumber\\
&V_4(x)=V_{+}(x,A-3)+2(A-1)+1+2(A-2)+1+2(A-3)+1\nonumber\\
\end{align}
These are plotted in figure 2. 
\begin{figure}
\centering
  \includegraphics[width = .5 \linewidth]{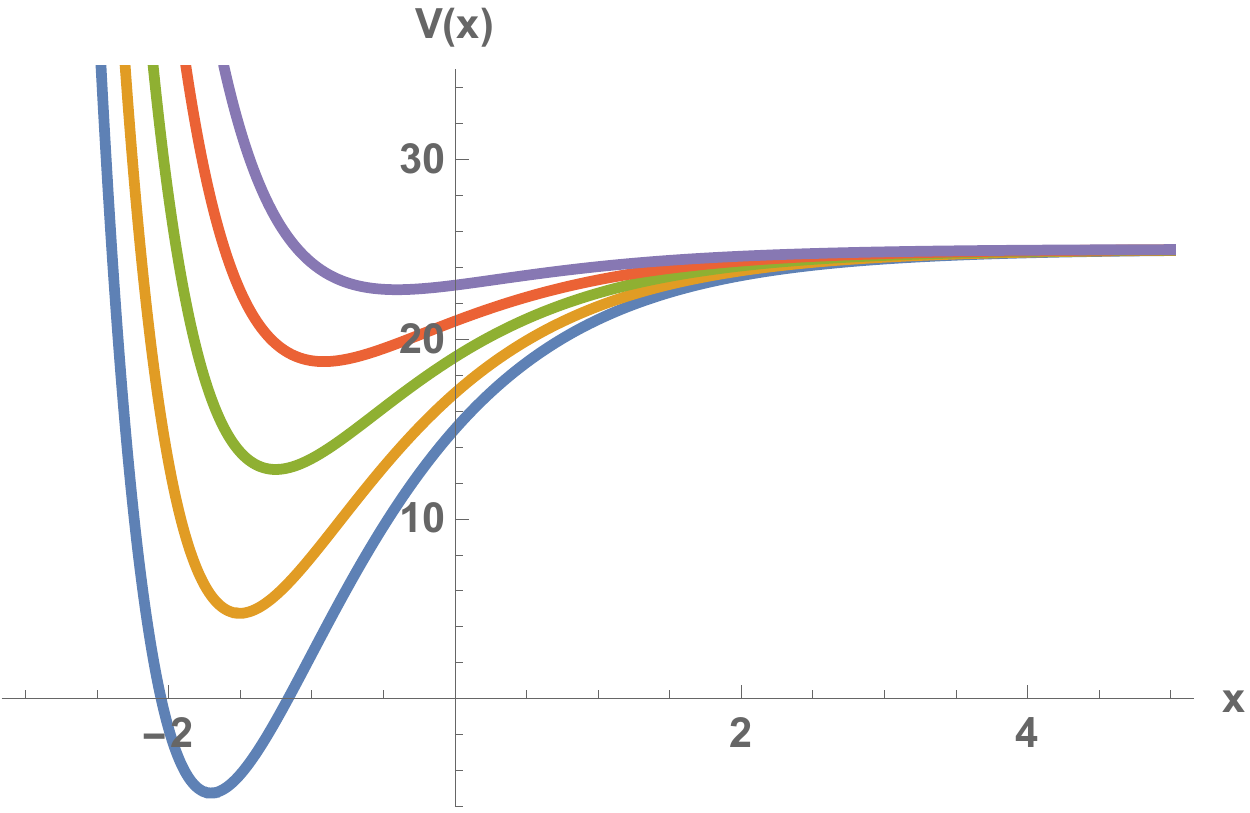}
  \caption{Hierarchy of potentials associated with the Morse potential for $A=5$.}
  \label{fig:Radion Potential}
\end{figure}
Exact solutions to the Schrodinger equation can be obtained using the ladder operators for the Morse potential. The expressions are given by:
with eigenfunctions:
\begin{equation}\psi _n^ - (x) = {e^{ - x(A - n)}}{e^{ - {e^{ - x}}}}L_n^{(2A - 2n)}(2{e^{ - x}})\end{equation}
with $L^{(k)}_n(y)$ the associated Laguerre polynomials. 
with bound state energies:
\begin{equation}E_n=A^2-(A-n)^2\end{equation}
For $A=5$ these are given by:
\begin{equation}\{0,9,16,21,24\}\end{equation}
which correspond to the ground state energies of the Hierarchy Hamiltonians in (2.9).

\section{VQE Calculation of the Ground State Energy in  Supersymmetric quantum mechanics}

In this section we describe the calculation of the ground state energy of the Morse potential in the supersymmetric quantum mechanics formulation of the Morse potential. Before one can set up the calculation to compute the ground state energies using the VQE one needs to perform a Hamiltonian mapping in terms of qubits. The first step is to represent the Hamiltonian as an $N \times N $ matrix using a discete quantum mechanics approximation to the quantum mechanical operators which would be infinite dimensional for bosonic observables \cite{Miceli}\cite{Okock}\cite{Korsch}\cite{Motycka}. In this paper we will use three different types of discrete Hamiltonians and compare the results from each.

\subsection*{Gaussian or Simple Harmonic Oscillator basis}

This is a very useful basis based on the matrix treatment of the simple harmonic oscillator which is sparse in representing the position and momentum operator. For the position operator we have:
\begin{equation} 
 X_{osc} = \frac{1}{\sqrt{2}}\begin{bmatrix}
 
   0 & {\sqrt 1 } & 0 &  \cdots  & 0  \\ 
   {\sqrt 1 } & 0 & {\sqrt 2 } &  \cdots  & 0  \\ 
   0 & {\sqrt 2 } &  \ddots  &  \ddots  & 0  \\ 
   0 & 0 &  \ddots  & 0 & {\sqrt {N-1} }  \\ 
   0 & 0 &  \cdots  & {\sqrt {N-1} } & 0  \\ 
\end{bmatrix}
  \end{equation}
while for the momentum operator we have:
\begin{equation}
 P_{osc} = \frac{i}{\sqrt{2}}\begin{bmatrix}
 
   0 & -{\sqrt 1 } & 0 &  \cdots  & 0  \\ 
   {\sqrt 1 } & 0 & -{\sqrt 2 } &  \cdots  & 0  \\ 
   0 & {\sqrt 2 } &  \ddots  &  \ddots  & 0  \\ 
   0 & 0 &  \ddots  & 0 & -{\sqrt {N-1} }  \\ 
   0 & 0 &  \cdots  & {\sqrt {N-1} } & 0  \\ 
\end{bmatrix}
  \end{equation}
The  Morse  Hamiltonian $H_{-}$ is then 
\begin{equation}
 H_{-}=P_{osc}^2 + Exp(-2X_{osc}) - (2A+1) Exp(-X_{osc}) + A^2 I   
\end{equation}
where $Exp$ refers to the Matrix exponential and $I$ is the $N \times N$ identity matrix.

\subsection*{Position basis}

In the position basis the position matrix is diagonal but the momentum matrix is dense and constructed from the position operator using a Sylvester matrix $F$. In the position basis the position matrix is:
\begin{equation}
{\left( {{X_{pos}}} \right)_{j,k}} = \sqrt {\frac{{2\pi }}{{4N}}} (2j - (N + 1)){\delta _{j,k}}
\end{equation}
and the momentum matrix is:
\begin{equation}{P_{pos}} = {F^\dag }{X_{pos}}F\end{equation}
where 
\begin{equation}{F_{j,k}} = \frac{1}{{\sqrt N }}{e^{\frac{{2\pi i}}{{4N}}(2j - (N + 1))(2k - (N + 1))}}\end{equation}
The Morse Hamiltonian is  formed from
\begin{equation}
 H_{-}=P_{pos}^2 + Exp(-2X_{pos}) - (2A+1) Exp(-X_{pos}) + A^2 I   
\end{equation}
but in this case the matrix exponential is very simple as it is the exponential of a diagonal matrix.

\subsection*{Finite difference basis}

This is the type of basis that comes up when realized differential equations in terms of finite difference equations. In this case the position operator is again diagonal but the momentum operator althogh not diagonal is still sparse. In the finite difference basis the position matrix is:
$${\left( {{X_{fd}}} \right)_{j,k}} = \sqrt {\frac{1}{{2N}}} (2j - (N + 1)){\delta _{j,k}}$$
and the momentum-squared matrix is:
\begin{equation} 
 P_{fd}^2 = \frac{N}{2}\begin{bmatrix}
 
   2 & - 1  & 0 &  \cdots  & 0  \\ 
   -1 & 2 & -1 &  \cdots  & 0  \\ 
   0 & -1 &  \ddots  &  \ddots  & 0  \\ 
   0 & 0 &  \ddots  & 2 & -1  \\ 
   0 & 0 &  \cdots  & -1 & 2  \\ 
\end{bmatrix}
  \end{equation}
The Morse Hamiltonian is then:
\begin{equation}
 H_{-}=P_{fd}^2 + Exp(-2X_{fd}) - (2A+1) Exp(-X_{fd}) + A^2 I   
\end{equation}
Whatever basis one uses one needs to map the Hamiltonian to a an expression in terms of a sum of tensor products of Pauli spin matrices plus the identity matrix which are called Pauli terms. As there are four such matrices the maximum number of terms in this expansion is $4^n$  where $n$ is the number of qubits, In most of our simulations the number of qubits was fixed at 4 so that the maximum number of Pauli terms was 256.

\subsection*{Calculation of bound state energies using the VQE for supersymmetric quantum mechanics}

\begin{figure}[ht]
\centering
\includegraphics[scale=0.4]{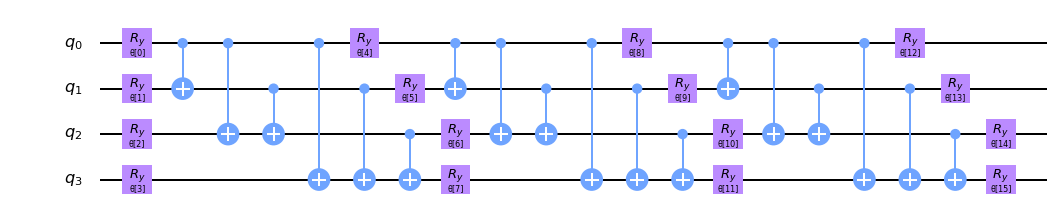}
\caption{A parameterized \(R_y\) variational form with 4 qubits, full entanglement and a depth of 3. This circuit is the quantum computing analog of an ansatz wavefunction.}
\label{fig:Ry}
\end{figure}

\begin{figure}[ht]
\centering
\includegraphics[scale=0.6]{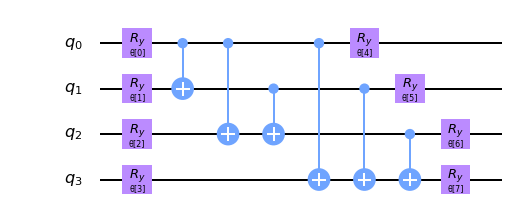}
\caption{One single unit (depth = 1) of an \(R_y\) variational form.}
\label{fig:RyDepth1}
\end{figure}

The VQE algorithm is a semi-quantum algorithm which is based on the variational method of quantum mechanics. The variational method allows one to calculate the upper bound of the ground state energy of a quantum system, without ever having to solve the Schrodinger equation. All that is needed is knowledge of the Hamiltonian, H, and for one to pick any ansatz wave function, \(\psi\). Mathematically, the variational method says that for this ansatz wave function:
\begin{equation}\label{variational}
E_0 \leq \braket{\psi|H|\psi} \equiv \langle H \rangle
\end{equation}
This means that the expectation value of the Hamiltonian, in the ansatz state, will always be an upper bound on the true ground state energy. If the ansatz is chosen well enough, the expectation value of the Hamiltonian can be made arbitrarily precise to the true ground state energy.
The VQE algorithm is an implementation of the variational method on a quantum computer. The idea is to create the ansatz wave function as a quantum circuit using parameterized quantum gates. A quantum circuit which represents the ansatz wave function is called the variational form. An example of a variational form is shown in Figure \ref{fig:Ry}, where a series of parameterized \(R_y\) gates are applied on each qubit, and each qubit is entangled to every other qubit using \(CNOT\) gates. This sort of entanglement is known as full entanglement. This circuit is known as an \(R_y\) variational form, where \(R_y\) represents a rotation of the qubit statevector about the \(y\)-axis on a Bloch sphere. The Hamiltonian is also mapped to a quantum circuit using the fundamental \(X\), \(Y\), \(Z\), and \(I\) gates. The VQE algorithm first sets arbitrary parameters for the variational form (in this case, arbitrary angles for each \(R_y\)) gate, then it calculates the expectation value of the Hamiltonian in the current ansatz state using a quantum computer. It then updates the parameters and calculates the expectation value again. Using a classical computer, it finds the cost function between the current and previous expectation value, after which it uses a classical optimizer to optimize the parameters until the expectation value is minimized. The efficiency and accuracy of results can be controlled by the depth, the entanglement, the variational form of the circuit, and optimizer used. The depth of the circuit represents how many times a variational form is repeated. For example, one unit of the \(R_y\) variational form applies an \(R_y\) gate to each qubit, then CNOT gates to entangle each qubit, then another set of \(R_y\) gates to each qubit. This is shown in Figure \ref{fig:RyDepth1}. A depth of 3, as shown in Figure \ref{fig:Ry}, repeats this pattern 3 times. A larger depth allows the variational form to generate a larger set of states; however, this comes at the cost of longer algorithm runtimes, as the number of parameters to be optimized is also increased.

\begin{figure}[ht]
\centering
\includegraphics[scale=0.3]{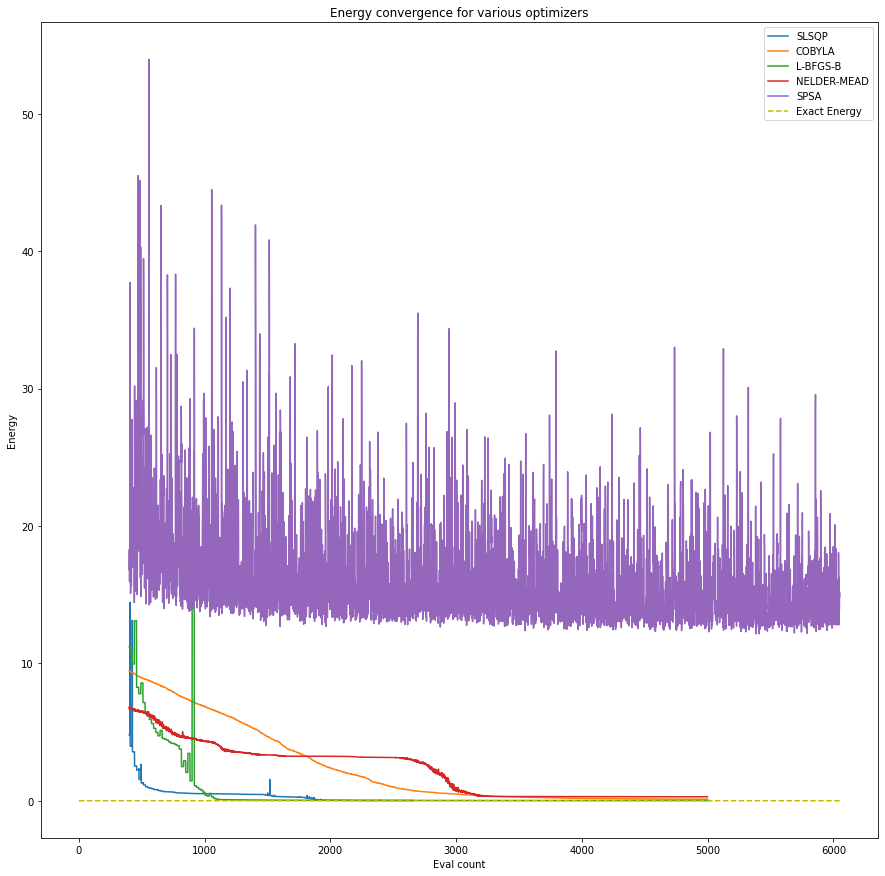}
\caption{Convergence plot for the various optimizers used in the VQE calculation of the Morse potential.}
\label{fig:optimizerConvergence}
\end{figure}

In calculating the bound state energies on the quantum computer it is convenient to scale the partner Hamiltonians by $1/2$. Then the $H_{-}$ and $H_{+}$ Hamiltonians for the Morse potential are written as:
\[{H_ - } = \frac{{{p^2}}}{2} + \frac{1}{2}\left( {{e^{ - 2x}} - (2A + 1){e^{ - x}} + {A^2}} \right)\]
\begin{equation}{H_ + } = \frac{{{p^2}}}{2} + \frac{1}{2}\left( {{e^{ - 2x}} - (2A - 1){e^{ - x}} + {A^2}} \right)\end{equation}
Performing a VQE calculation using IBM QISKit, we find accurate results for the case $A=5$ given by $E_0 = 0.0137$ for $H_{-}$ and $E_1=4.5428$ for $H_{+}$ in the oscillator basis. This compares well with the exact values of $E_0=0$ and $E_1= 9/2$ for $A=5$. These results are shown in Table \ref{tab:BasisCompare}. The convergence plots for these calculations are shown in Figure 5s. The VQE algorithm was then run on the \(H_-\) Hamiltonian using different optimizers. The result for each optimizer is shown in Table \ref{tab:optimizerCompare}. The convergence plot for the various optimizers is shown in Figure \ref{fig:optimizerConvergence}. Note that is all cases the VQE result lies above the exact energy value. This is because the variational method provides a lower bound of the energy $E_{VQE} \ge E_0$. 
\\
\begin{table}[ht]
\centering
\begin{tabular}{|l|l|l|l|}
\hline
Basis       & Hamiltonian  & VQE Result & No. Pauli Terms \\ \hline
Oscillator   & \(H_-\) & 0.0137  & 135 \\ \hline
Oscillator   & \(H_+\) & 4.5428  & 135 \\ \hline

\end{tabular}
\caption{\label{tab:BasisCompare}  VQE results for Morse potential with $A=5$ using the oscillator basis. All of the Hamiltonian were mapped to 4-qubit operators. The quantum circuit for each simulation utilized an \(R_y\) variational form, with a fully entangled circuit of depth 3. The backend used was a statevector simulator. The Sequential Least SQuares Programming (SLSQP) optimizer was used, with a maximum of 600 iterations. The exact result for $H_{-}$ and $H_{+}$ for the Morse potential used was $0$ and $4.5$ respectively.}
\end{table}

\begin{table}[ht]
\centering
\begin{tabular}{|l|l|l|l|}
\hline
Optimizer & VQE Result\\ \hline

SLSQP & 0.0182 \\ \hline
COBYLA & 0.1114\\ \hline
L-BFGS-B & 0.0123 \\ \hline
NELDER-MEAD & 0.2909\\ \hline
SPSA & 12.5363 \\ \hline

\end{tabular}
\caption{\label{tab:optimizerCompare}  VQE results for \(H_-\) obtained using various optimizers.}
\end{table}

Finally it is possible by re-scaling the $x$ coordinate to write the Hamiltonion $H_{-}$ in the form
\begin{equation}{H_ - } = \frac{{{p^2}}}{2} + \frac{1}{2}{\left( {A + \frac{1}{2}} \right)^2}\left( {{e^{ - 2x}} - 2{e^{ - x}} + 1} \right) - \frac{1}{2}\left( {A + \frac{1}{4}} \right)\end{equation}
This form of the Morse potential similar to that used to describe diatomic molecules which we discuss in the next section.

\begin{figure}[!htb]
\centering
\minipage{0.5\textwidth}
  \includegraphics[width=\linewidth]{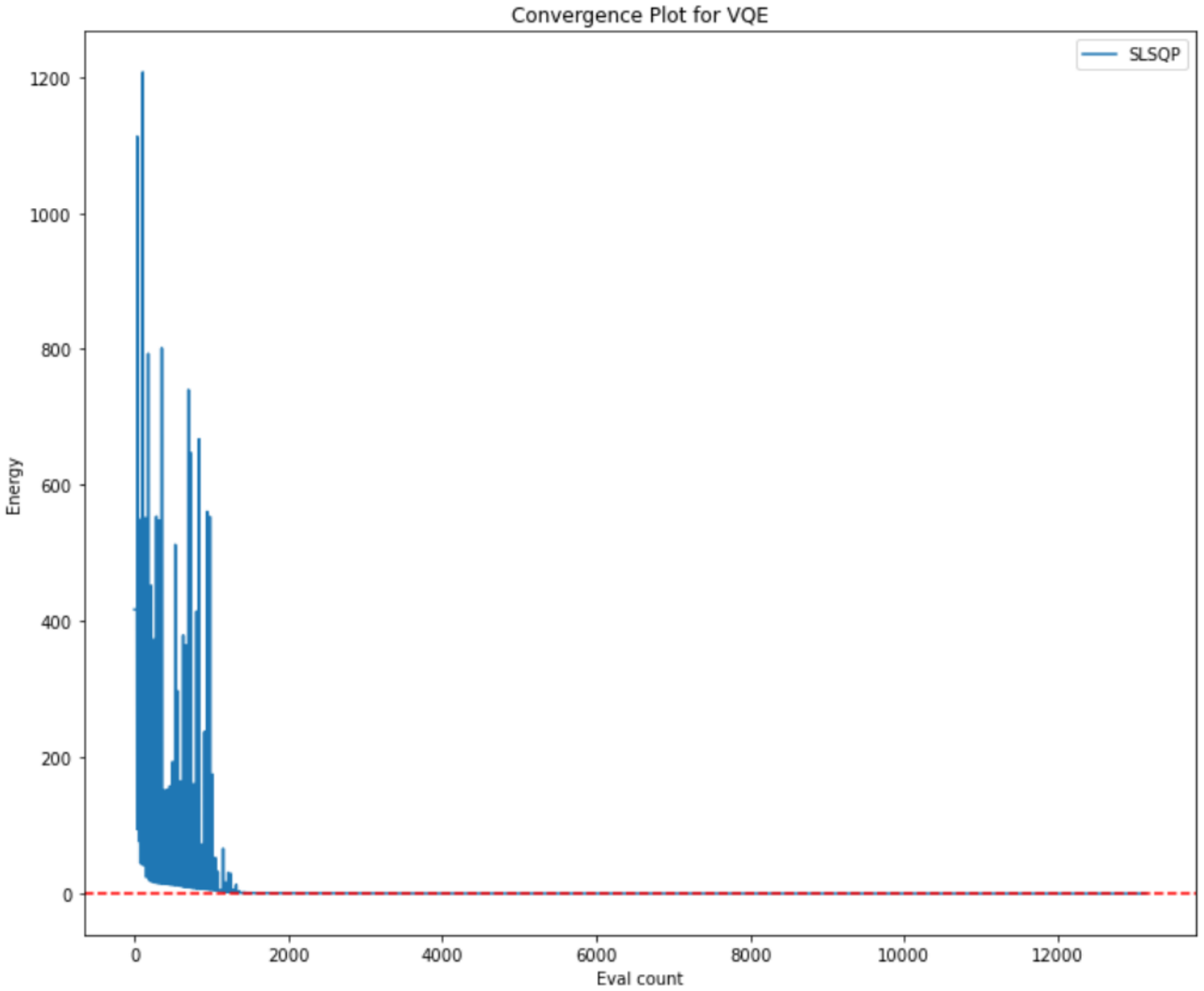}
\endminipage\hfill
\minipage{0.5\textwidth}
  \includegraphics[width=\linewidth]{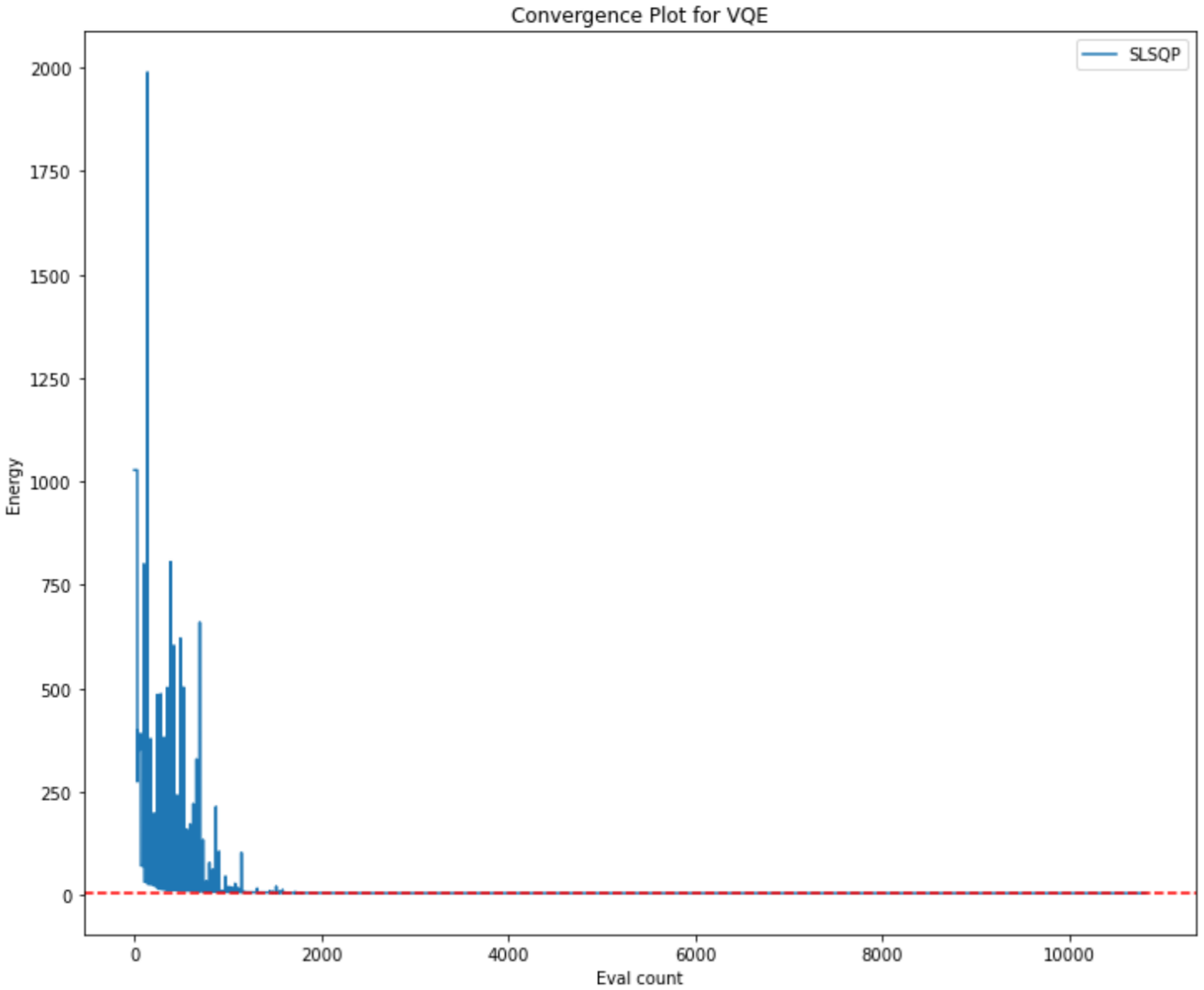}
\endminipage\hfill
\caption{(Left) Convergence plot for the VQE for the minus partner Hamiltonian (Right) Convergence plot for the VQE for the plus partner Hamiltonian using the oscillator basis.}
\end{figure}

\section{Calculation of bound state energies using the VQE for realistic diatomic molecules}

The Hamiltonian for the Morse potential describing diatomic molecules is given by \cite{Fidiani}:
\begin{equation}{H_M} = \frac{{{p^2}}}{{2{m_r}}} + {V_M}(x)\end{equation}
where the Morse potential is given by:
\begin{equation}{V_M} = D_{mol}({e^{ - 2a x}} - 2{e^{ -a x}})\end{equation}
with the reduced mass of the diatomic molecule given. by:
\begin{equation}{m_r} = \frac{{{m_1}{m_2}}}{{{m_1} + {m_2}}}\end{equation}
To study the Morse potential on a quantum computer it is more convenient to study the Hamiltonian
\begin{equation}H = \frac{{{p^2}}}{2} + \frac{{{\lambda ^2}}}{2}({e^{ - 2x}} - 2{e^{ - x}} + 1)\end{equation}
This is related to the Morse Hamiltonian in a simple way.
If the ground state energy of this Hamiltonian is $\varepsilon_0$ the ground state energy of the molecule is
\begin{equation}{E_0} = ({\varepsilon _0} - \frac{{{\lambda ^2}}}{2}){E_{mult}}\end{equation}
where $\lambda^2$ and $E_{mult}$ as well as other parameters are listed for various molecules in Tables 2, 3 and 4.
In terms of $\varepsilon_0$ we have 
\begin{equation}{\varepsilon _0} = \frac{{\lambda  - \frac{1}{4}}}{2}\end{equation}
and the potential asymptotes to the constant value at $x \to \infty $
\begin{equation}{\varepsilon _\infty } = \frac{{{\lambda ^2}}}{2}\end{equation}
and the ratio:
\begin{equation}\frac{{{\varepsilon _0}}}{{{\varepsilon _\infty }}} = \frac{1}{\lambda } - \frac{1}{{4{\lambda ^2}}}\end{equation}
In terms of the parameters of the molecule
\[{\lambda ^2} = \frac{{2{m_r c^2}{D_{mol}}}}{{{a^2}{\hbar ^2}{c^2}}}\]
\begin{equation}{E_{mult}} = \frac{{{a^2}{\hbar ^2}{c^2}}}{{{m_r c^2}}}\end{equation}
We plot some of the potentials in figures 7 and 8. Tables 3, 4 and 3  contain the results for the VQE calculations using 4 qubits and the the oscillator basis. In all cases we found excellent agreement with the experimentally measured values. Convergence plots for the VQE calculations are given in figures 9 and 10 and all cases rapidly converge near the ground state energy.
\newpage
\begin{table}[h]
\centering
\begin{tabular}{|l|l|l|l|l|l|}
\hline
Molecule       & $\lambda^2$  & $E_{mult}$ & No. bound states & $E_0 (eV)$ & $E_0 (eV)$ VQE \\ \hline
$H_2$   &    303.128         & 0.0313043   & 17   & -4.476 & -4.468 \\ \hline
$HCl $            & 619.618        & 0.014906           & 25    &   -4.43434 &    -4.43222         \\ \hline
$LiH$          & 832.439                    &0.00604318  & 29     & -2.42886 &      -2.42807          \\ \hline
$CO$            & 6969.27                   & 0.00322146   & 83  & -11.0915 &   -11.0911                \\ \hline
$O_2$            & 2831.21                  & 0.00368323 & 53  & -5.11647  &    -5.11599                \\ \hline
$N_2$            & 4581.05                   & 0.00432434   & 68  & -9.7592 &    -9.75869                \\ \hline

\end{tabular}
\caption{\label{tab:table-name} Various  parameters used in the calculation of the ground state energy for the Morse potential for diatomic molecules. In all cases the IBM QISKit Variational Quantum Eigensolver (VQE) produced an accurate computation for the ground state energy.}
\end{table}
\begin{table}[h]
\centering
\begin{tabular}{|l|l|l|l|l|l|}
\hline
Molecule       & $\lambda^2/2$  &$\lambda$  & $\varepsilon_0/\varepsilon_\infty$ &$\varepsilon_0$  & $\varepsilon_0$ VQE  \\ \hline
$H_2$   &    151.564         & 17.4106   & 0.0566117   &8.58028 &  8.83606\\ \hline
$HCl $            & 309.809        & 24.8921          & 0.0397699    &   12.3211 &   12.4643
          \\ \hline
$LiH$          & 416.219                    &28.852  & 0.0343593     & 14.301 &     14.433          \\ \hline
$CO$            & 3484.63                  &83.4821    & 0.0119427  & 41.6161 &     41.7399               \\ \hline
$O_2$            &1415.6                   & 53.2091   & 0.0187055  & 26.4796 &    26.6034              \\ \hline
$N_2$            & 2290.53                   & 67.6835   &0.0147201  & 33.7167 &     33.8406               \\ \hline

\end{tabular}
\caption{\label{tab:table-name} Various  parameters used in the calculation of the ground state energy for the Morse potential for diatomic molecules using re-scaled variables and energies. In all cases the IBM QISKit Variational Quantum Eigensolver (VQE) produced an accurate computation for the ground state energy.}
\end{table}
\begin{table}[h]
\centering
\begin{tabular}{|l|l|l|l|l|l|}
\hline
Molecule       & $m_r (amu)$  & $D_{mol} (eV)$ & $a (10^{10} m^{-1})$  \\ \hline
$H_2$   &    .50391         & 4.7446   & 1.9426    \\ \hline
$HCl $            & .9796        & 4.618           & 1.869        \\ \hline
$LiH$          & 0.8801221                    &2.515287  & 1.1280         \\ \hline
$CO$            & 6.8606719                  & 11.2256  & 2.2994                 \\ \hline
$O_2$            & 8                  & 5.214 & 2.655                \\ \hline
$N_2$            & 7                  & 9.905  & 2.691              \\ \hline

\end{tabular}
\caption{\label{tab:table-name} Various parameters used in the computation of the ground state energies for the Morse potential for diatomic molecules.}
\end{table}

\newpage
\begin{figure}[!htb]
\centering
\minipage{0.5\textwidth}
  \includegraphics[width=\linewidth]{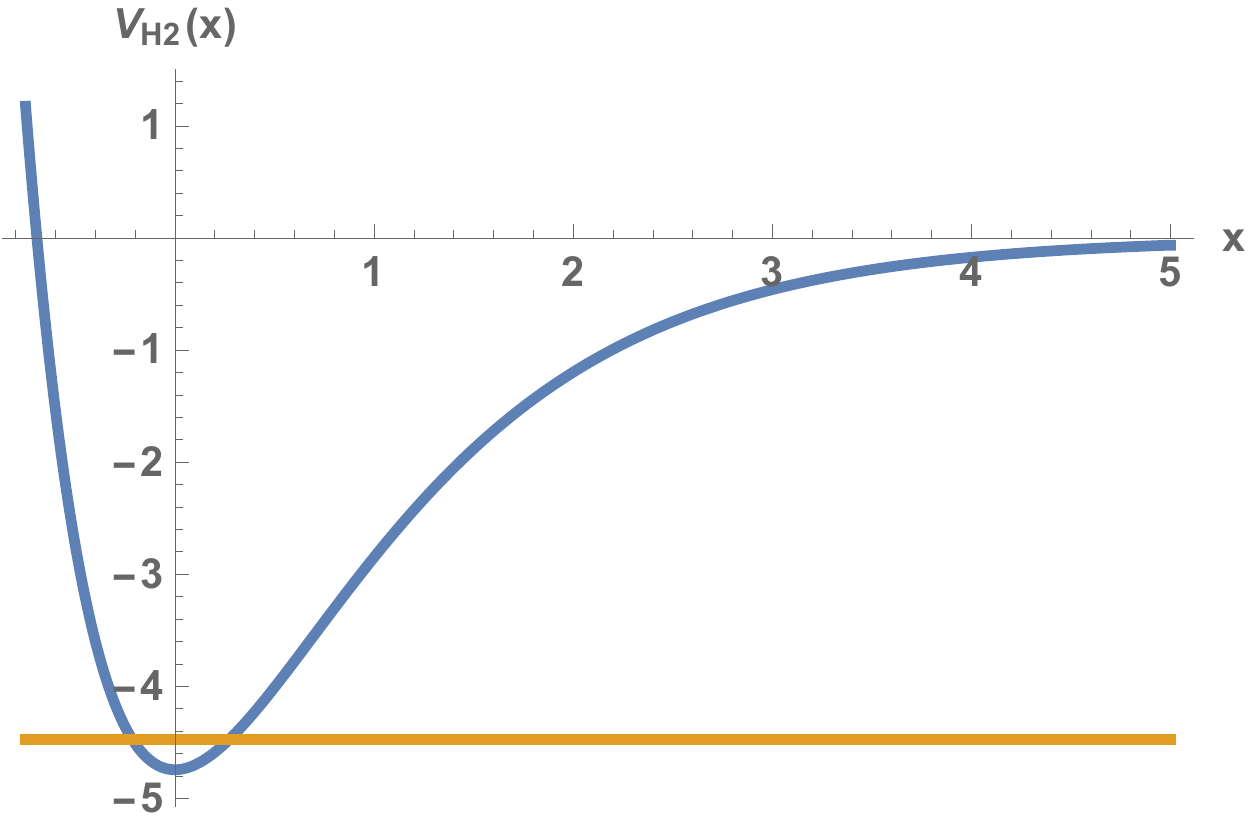}
\endminipage\hfill
\minipage{0.5\textwidth}
  \includegraphics[width=\linewidth]{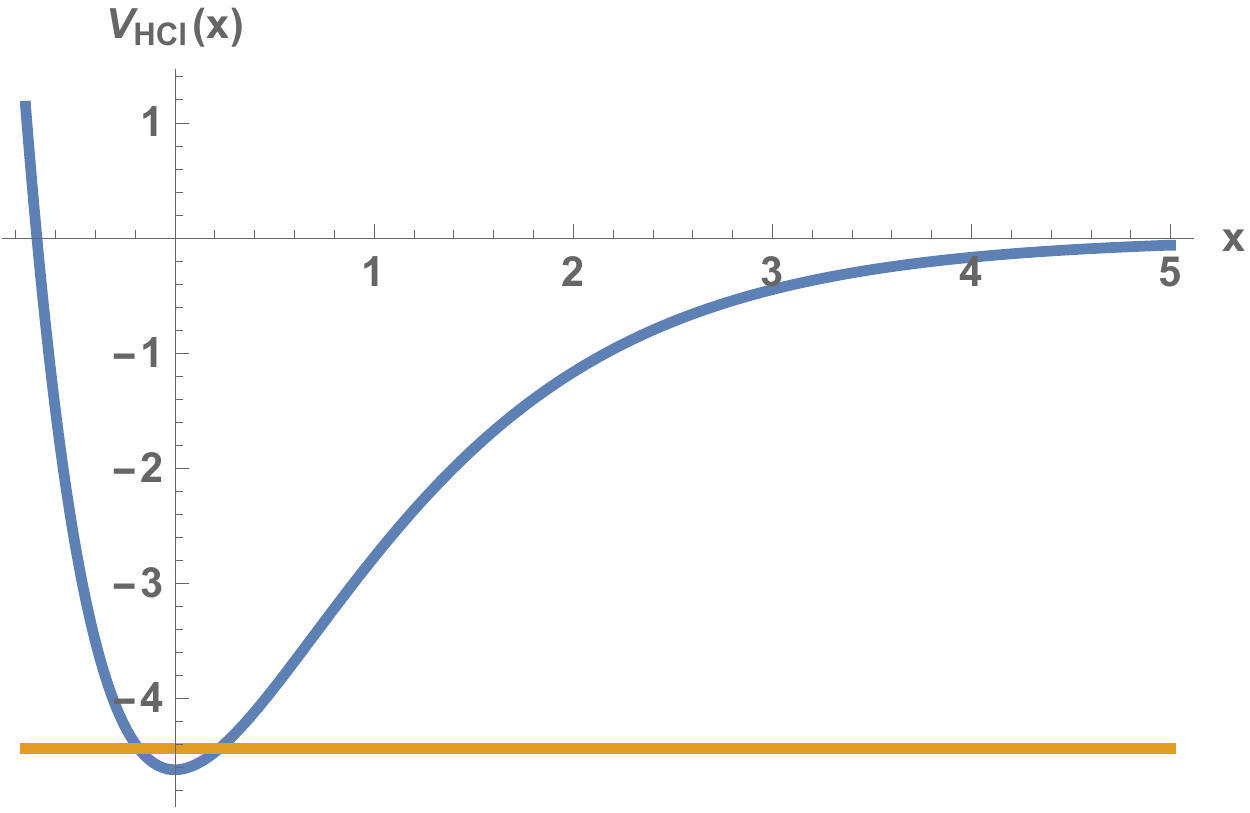}
\endminipage\hfill
\caption{(Left) Morse potential for the $H_2$ molecule (Right) Morse potential for the $HCl$ molecule.}
\end{figure}
\begin{figure}[!htb]
\centering
\minipage{0.5\textwidth}
  \includegraphics[width=\linewidth]{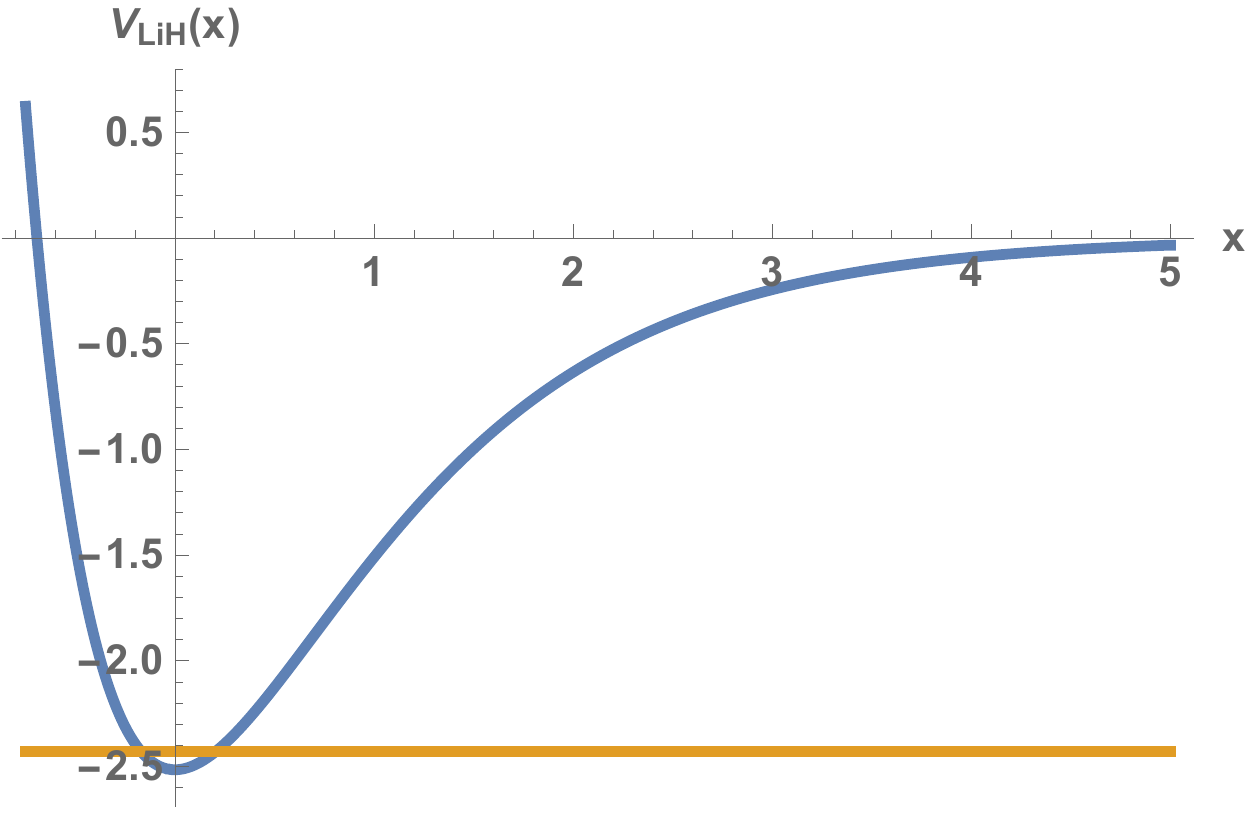}
\endminipage\hfill
\minipage{0.5\textwidth}
  \includegraphics[width=\linewidth]{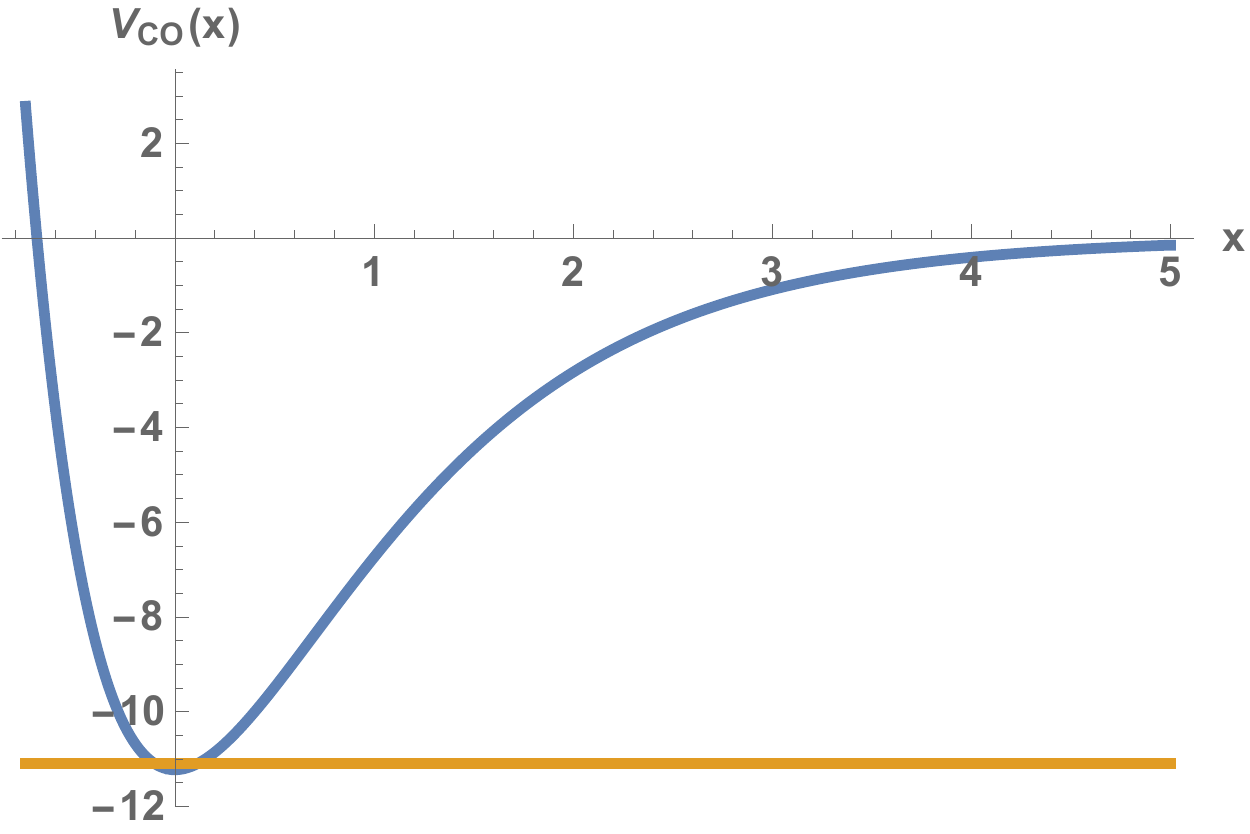}
\endminipage\hfill
\caption{(Left) Morse potential for the $LiH$ molecule. (Right) Morse potential for the $CO$ potential.}
\end{figure}
\begin{figure}[!htb]
\centering
\minipage{0.32\textwidth}
  \includegraphics[width=\linewidth]{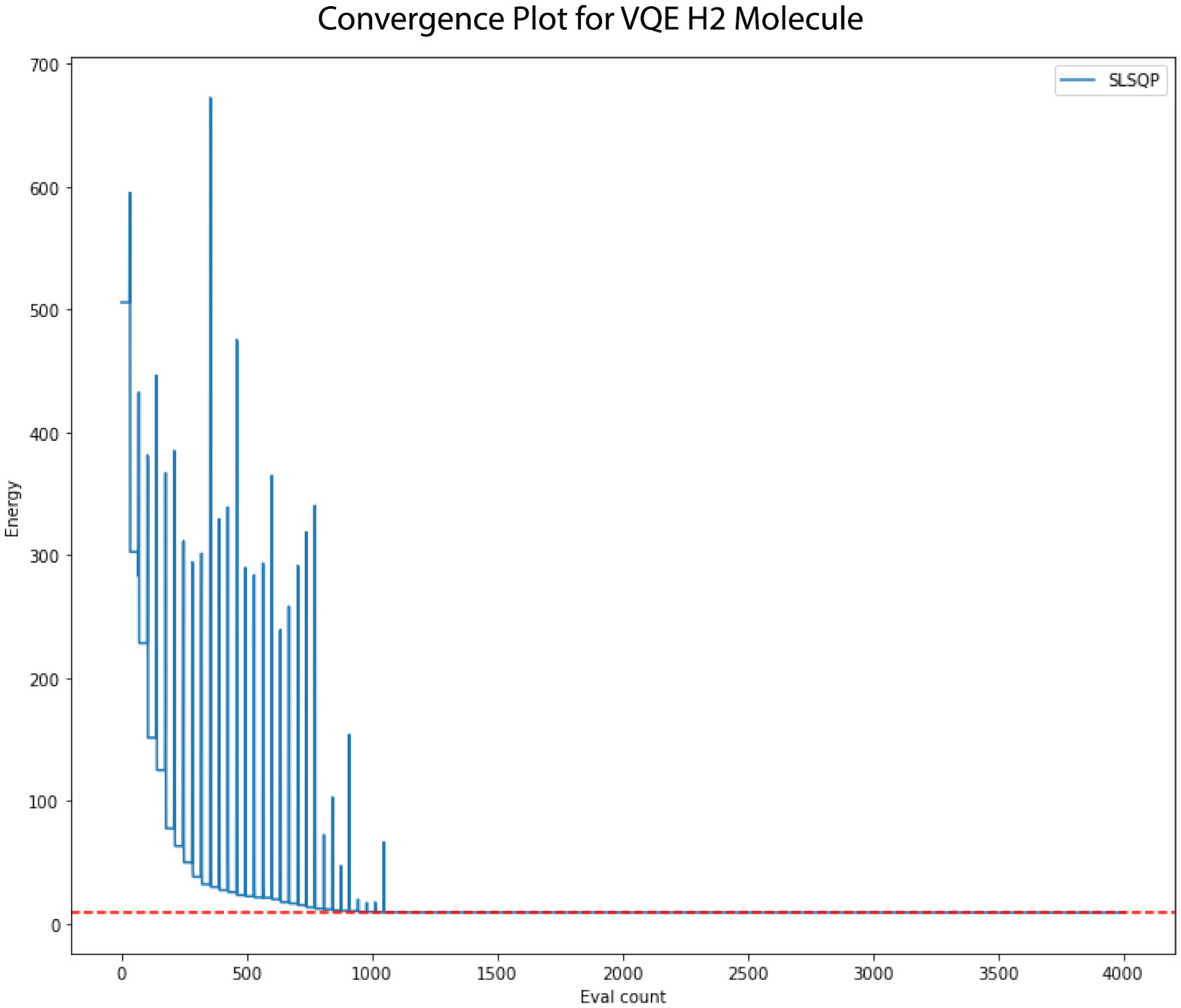}
\endminipage\hfill
\minipage{0.32\textwidth}
  \includegraphics[width=\linewidth]{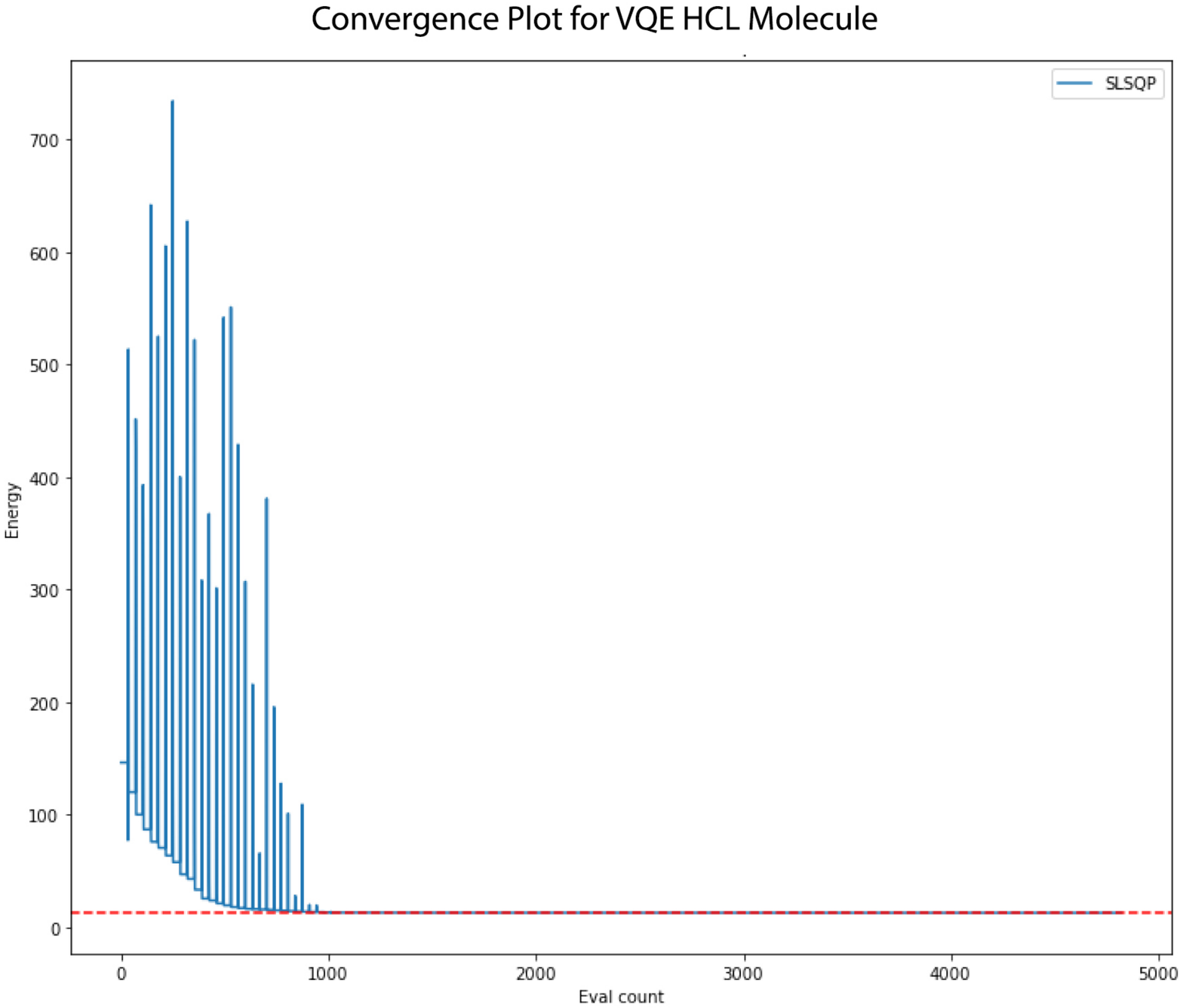}
\endminipage\hfill
\minipage{0.32\textwidth}%
  \includegraphics[width=\linewidth]{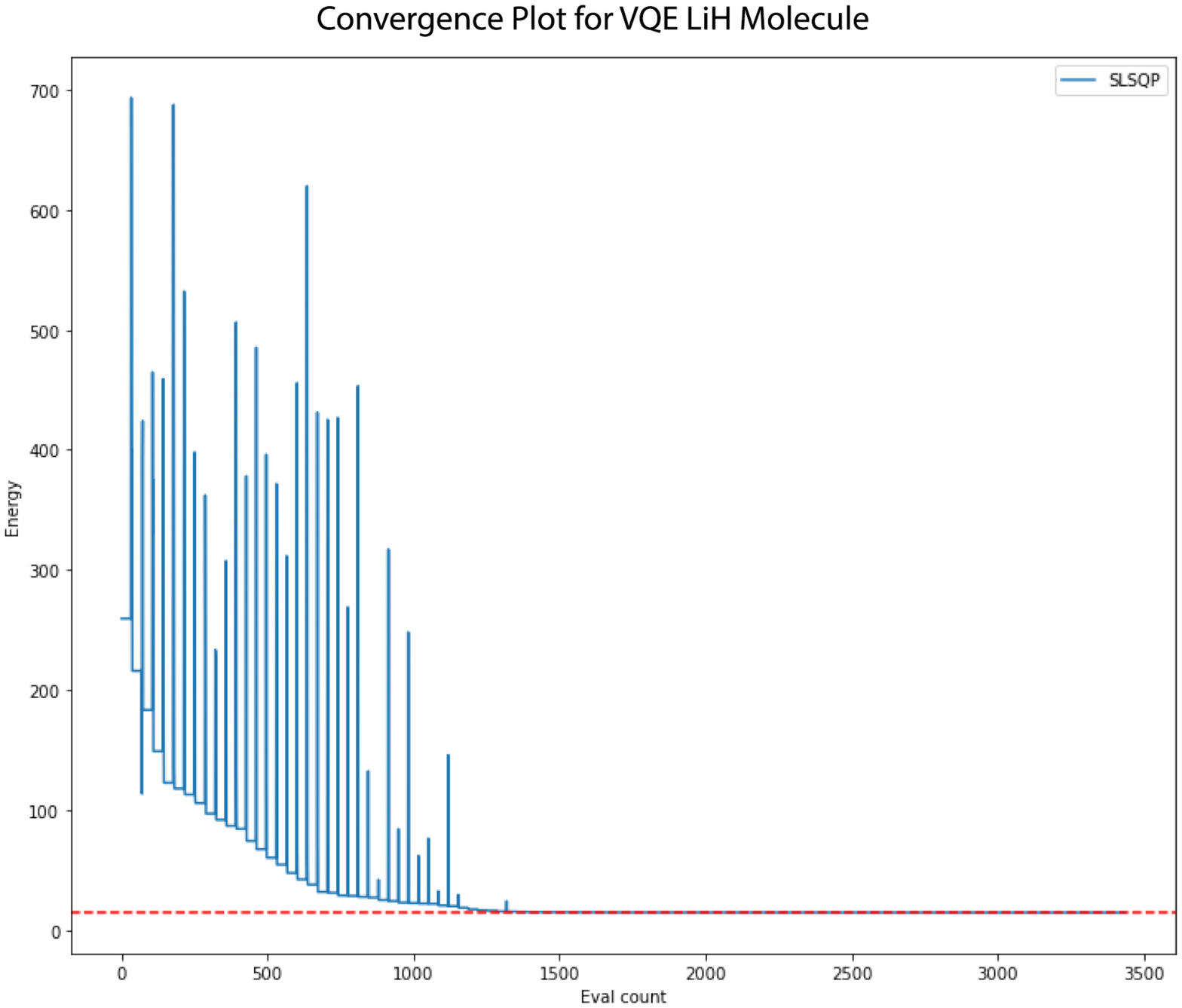}
 
\endminipage
\caption{(Left) VQE convergence plot for $H_2$ versus the number of optimization steps.(Middle) VQE convergence plot for $HCl$ versus the number of optimization steps. (Right) VQE convergence plot for $LiH$ versus the number of optimization steps.}
\end{figure}
\begin{figure}[!htb]
\centering
\minipage{0.32\textwidth}
  \includegraphics[width=\linewidth]{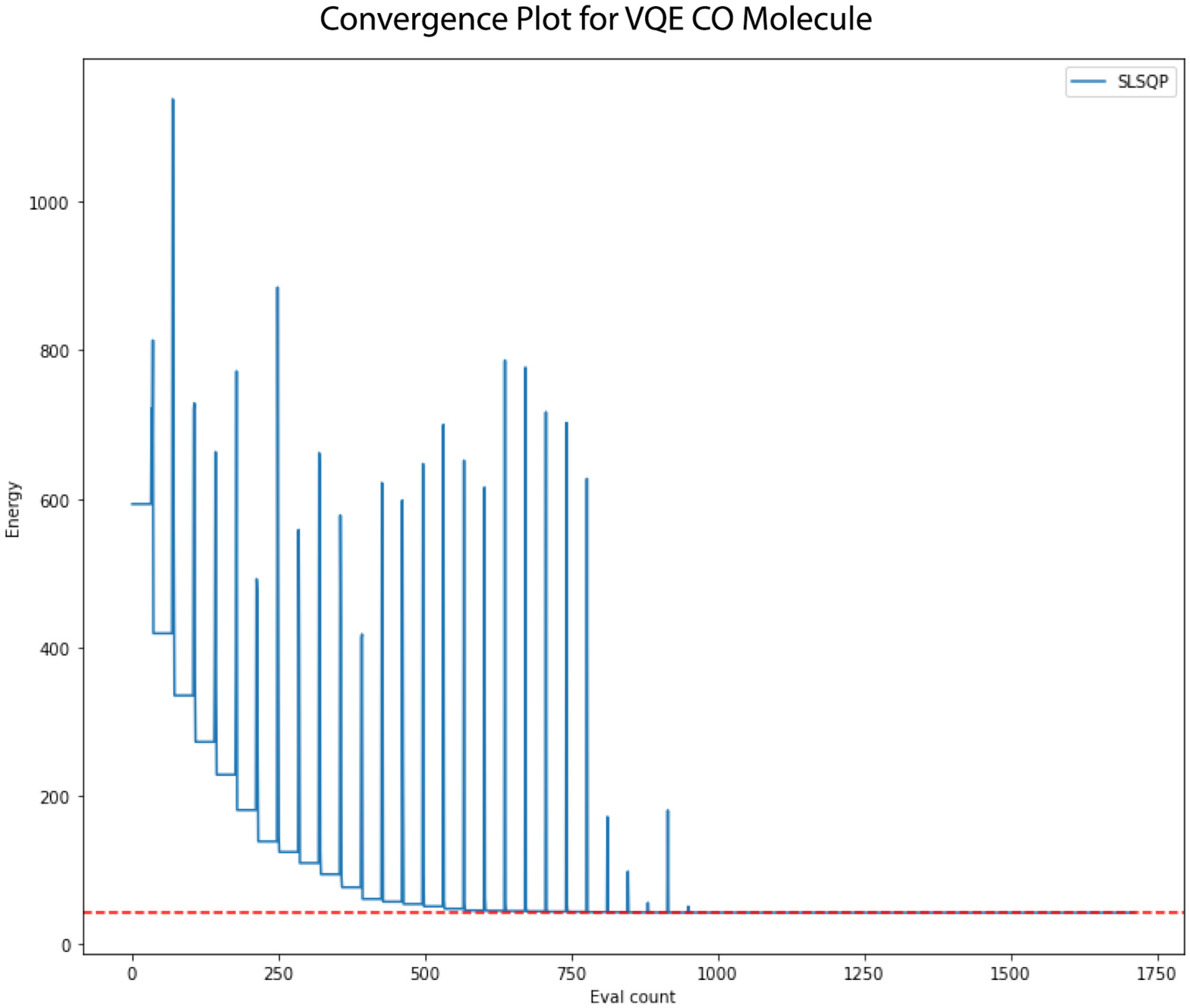}
\endminipage\hfill
\minipage{0.32\textwidth}
  \includegraphics[width=\linewidth]{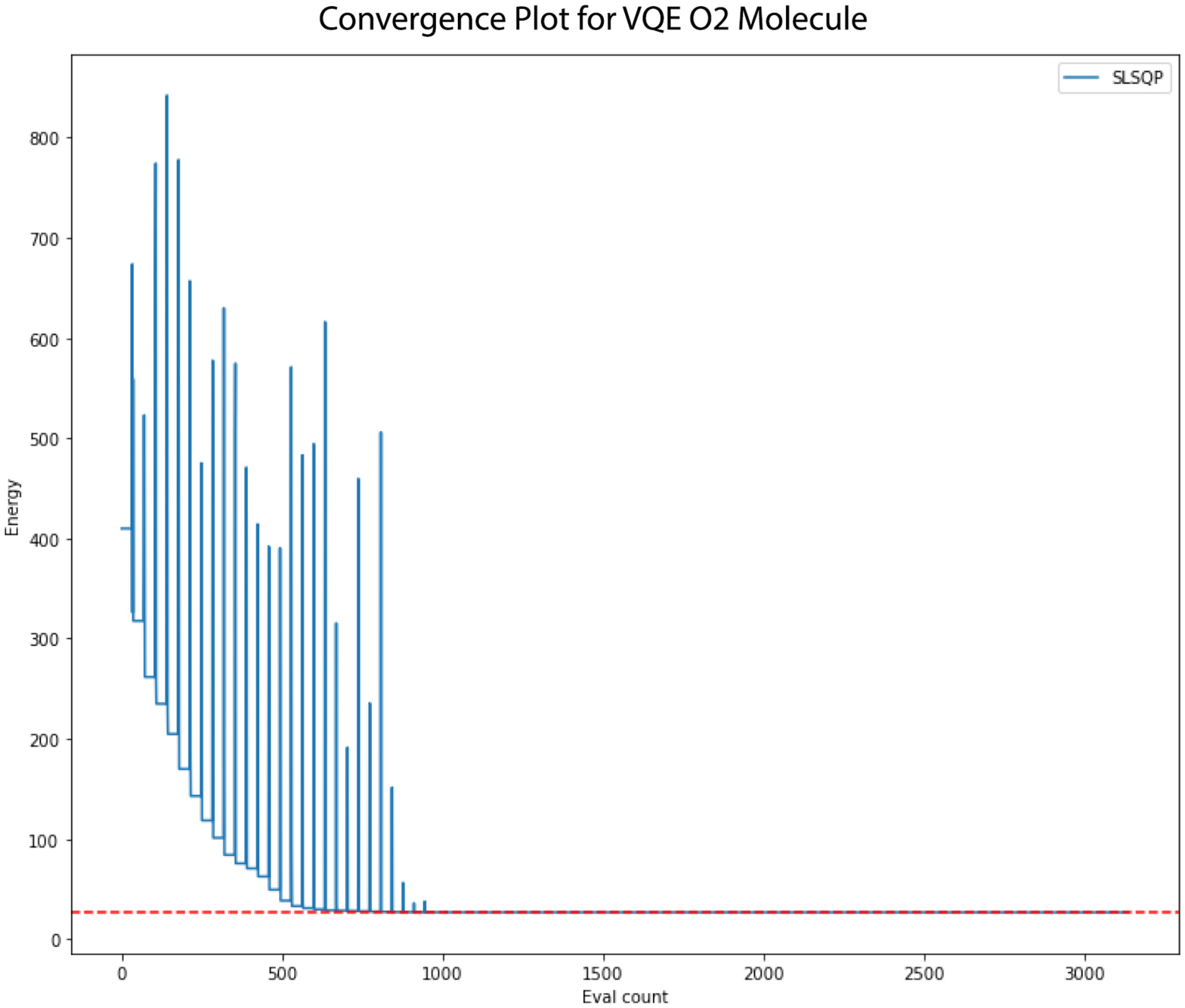}
\endminipage\hfill
\minipage{0.32\textwidth}%
  \includegraphics[width=\linewidth]{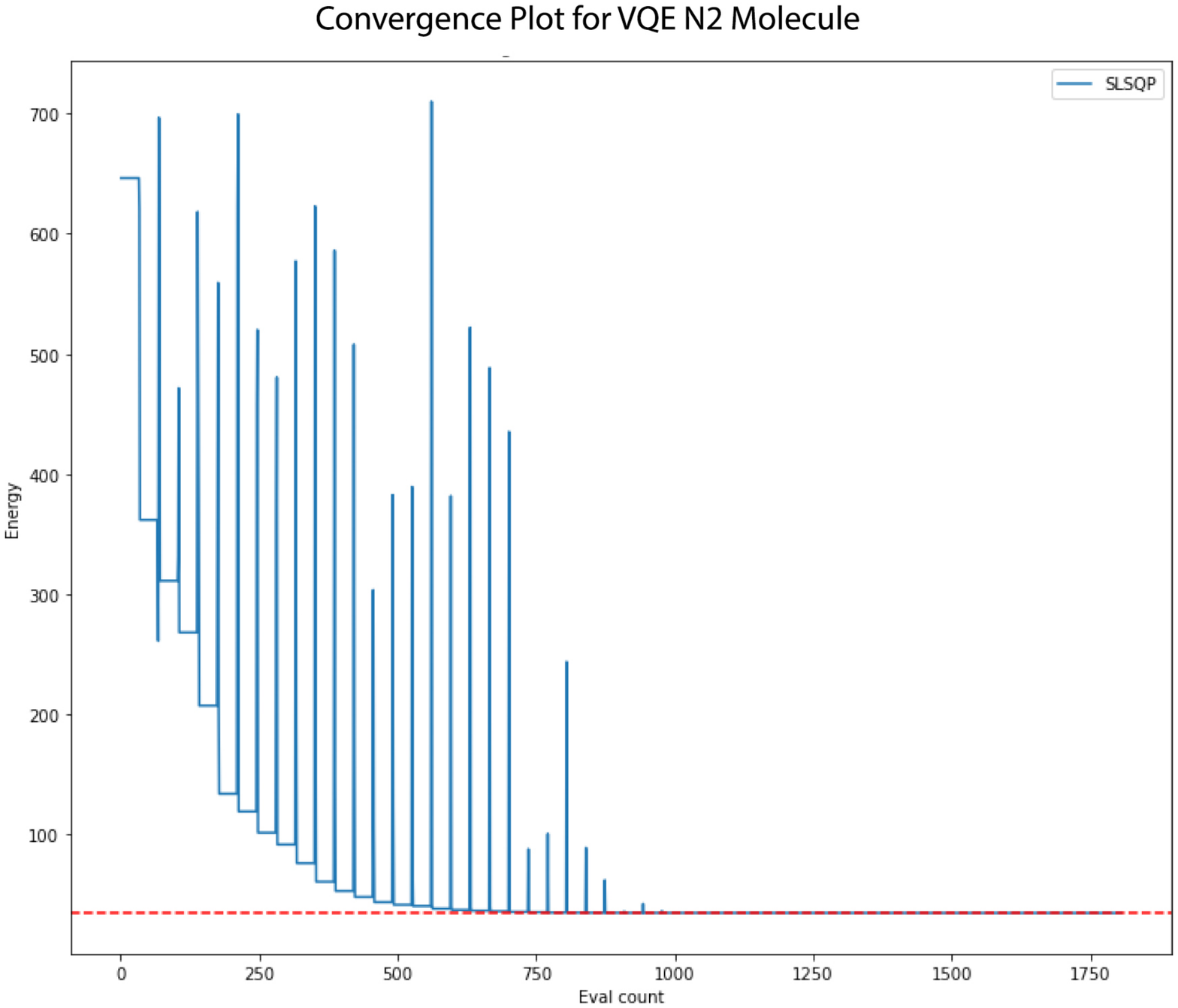}
\endminipage
\caption{(Left) VQE convergence plot for $CO$ versus the number of optimization steps.(Middle) VQE convergence plot for $O_2$ versus the number of optimization steps. (Right) VQE convergence plot for $N_2$ versus the number of optimization steps.}
\end{figure}
\newpage
\section{Morse Potential for Triatomic Molecules}

\begin{figure}[ht]
\centering
\includegraphics[scale=0.122]{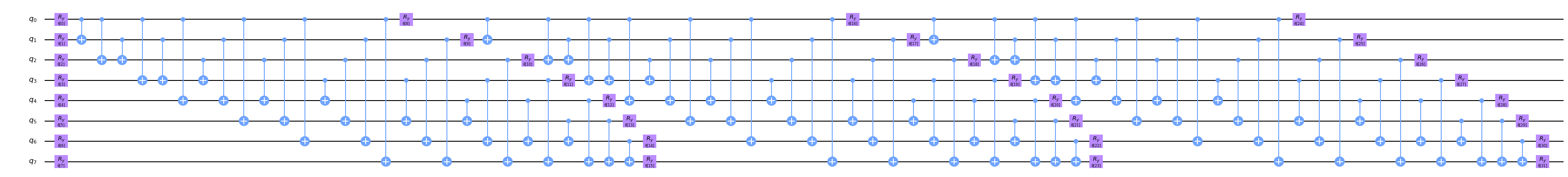}
\caption{8-qubit \(R_y\) variational form of depth 3 used in the VQE calculation of the Morse potential for triatomic molecules.}
\label{fig:2DMorse_varForm}
\end{figure}

The Morse potential can be extended to treat triatomic molecules as well \cite{Bordoni}\cite{Chesick}. In this case the Morse potential depends on two coordinates $x, y$ and Hamiltonian can be written as :
\begin{equation}\label{H1}{H_1} = \frac{{p_x^2}}{{2m}} + \frac{{p_y^2}}{{2m}} - \frac{{{p_x}{p_y}}}{M} + C{\left( {1 - {e^{ - x/b}}} \right)^2} + C{\left( {1 - {e^{ - y/b}}} \right)^2}\end{equation}
Another form of the Hamiltonian can be formed by changing variables
\[x = ({x_1} + {x_2})/\sqrt 2 \]
\begin{equation}y = ({x_1} - {x_2})/\sqrt 2 \end{equation}
and defining 
\[{m_1} = \frac{m}{{1 - \frac{m}{M}}}\]
\begin{equation}{m_2} = \frac{m}{{1 + \frac{m}{M}}}\end{equation}
In this form of the Hamiltonian can be written:
\begin{equation}\label{H2}{H_2} = \frac{{p_1^2}}{{2{m_1}}} + \frac{{p_2^2}}{{2{m_2}}} + C{\left( {1 - {e^{ - \left( {{x_1} + {x_2}} \right)/\sqrt 2 b}}} \right)^2} + C{\left( {1 - {e^{ - \left( {{x_1} - {x_2}} \right)/\sqrt 2 b}}} \right)^2}\end{equation}
A plot of this potential is given in Figure 12. For the case $m=1$, $M=2$, $C= 10$ and $b=\sqrt{20}$, we find the ground state energy $E_0 = 0.954585$ using either Hamiltonian.  Defining the tensor product operators 
\[{X_1} = {X_{osc}} \otimes I\]
\[{X_2} = I \otimes {X_{osc}}\]
\[{P_1} = {P_{osc}} \otimes I\]
\begin{equation}{P_2} = I \otimes {P_{osc}}\end{equation}
and using 8 qubits for the definition of the Hamiltonian in Equation \ref{H1}, we find $E_0= 0.957527$ using the VQE algorithm. A table of the simulation parameters and results for both \(H_1\) and \(H_2\) are presented in Table \ref{tab:TriatomicVQE}. The variational form for both simulations is shown in Figure \ref{fig:2DMorse_varForm}.

\begin{table}[h]
\centering
\begin{tabular}{|l|l|l|l|}
\hline
Hamiltonian & VQE Result & No. Pauli Terms \\ \hline
\(H_1\) & 0.957527 & 1293 \\ \hline
\(H_2\) & 0.959272  & 8808 \\ \hline
\end{tabular}
\caption{\label{tab:TriatomicVQE} VQE results for triatomic Morse potential for the Hamiltonians \(H_1\) and \(H_2\). The parameters in each equation were set as $m=1$, $M=2$, $C= 10$ and $b=\sqrt{20}$. Both Hamiltonians were mapped to an 8-qubit operator. The quantum circuit for each simulation utilized an \(R_y\) variational form, with a fully entangled circuit of depth 3. The backend used was a statevector simulator. The SLSQP optimizer was used, with a maximum of 600 iterations. The exact ground state energy for both Hamiltonians is $E_0 = 0.954585$.}
\end{table}

\begin{figure}[!htb]
\centering
\minipage{0.5\textwidth}
  \includegraphics[width=\linewidth]{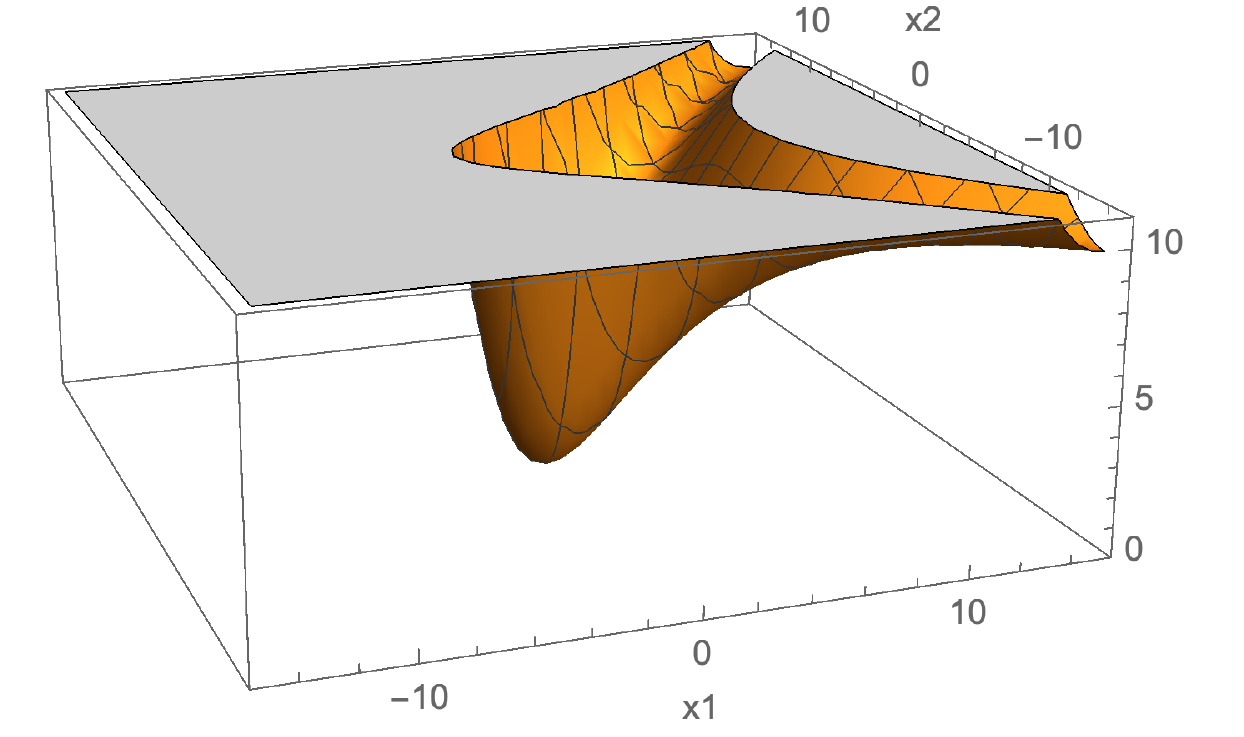}
\endminipage\hfill
\minipage{0.5\textwidth}
  \includegraphics[width=\linewidth]{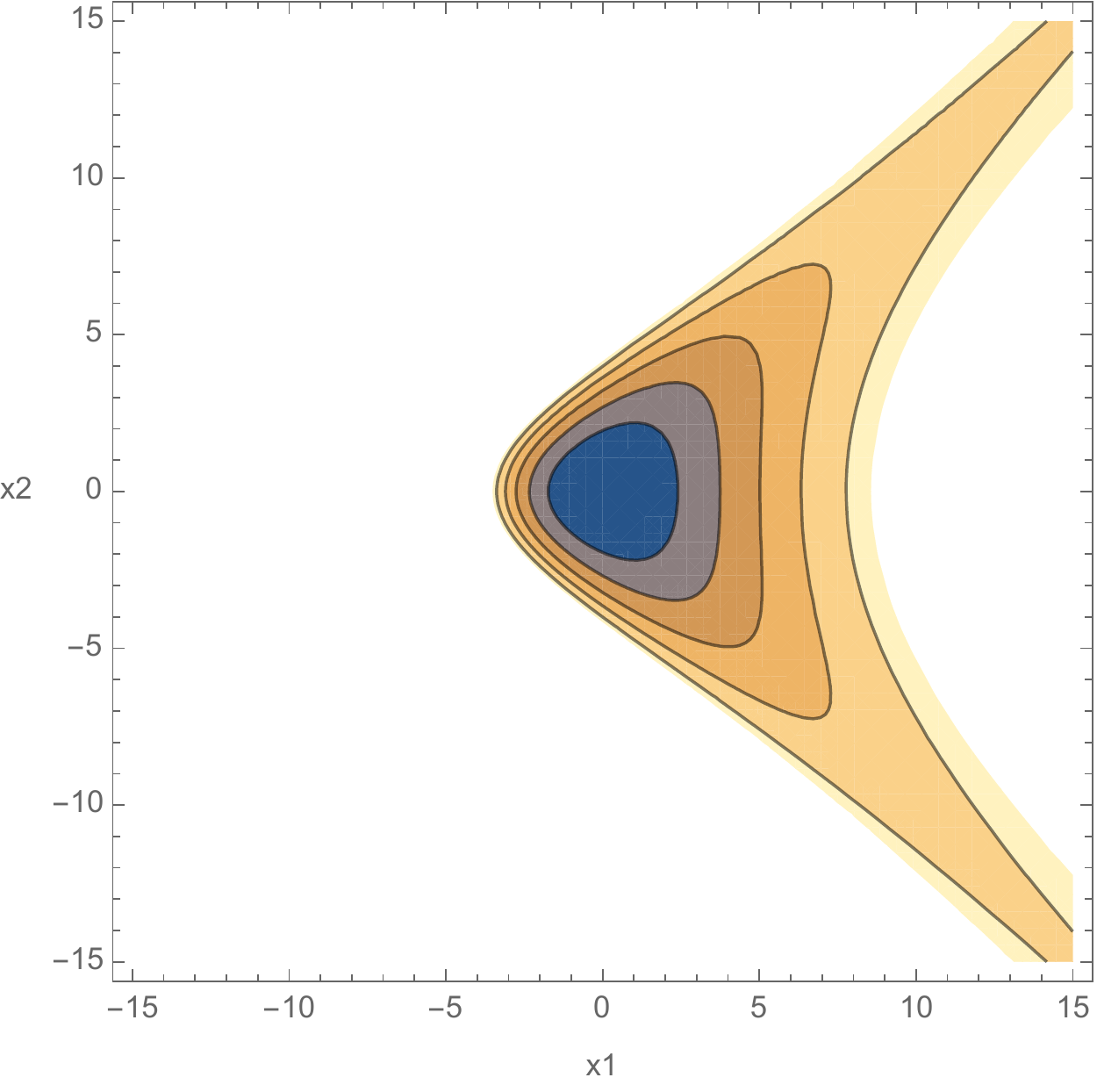}
\endminipage\hfill
\caption{(Left) 3D plot of a form of the Morse potential that can be applied to  triatomic molecules (Right) Contour plot of a form of the Morse potential that can be applied to triatomic  molecules.}
\end{figure}

\newpage
\section{Conclusion}

In this paper we investigated using a quantum computer to simulate Hamiltonians using the Morse potential for supersymmetric quantum mechanics, diatomic molecules and triatomic molecules. In all cases we found excellent agreement with the exact answers using IBM QISKit and its statevector simulator. It will be interesting to extend these calculations to other molecules and combine these methods with electronic structure calculations on near term quantum computers which can determine the parameters of the Morse potential \cite{Armaos}\cite{Choo}. This approach was used in \cite{Stober} to compute thermodynamic observables of the molecular systems. Finally it is important to study other variational forms as well as error and noise mitigation on near term devices associated with the calculations of the Morse potential to gain further understanding of the applicability of this effective Hamiltonian approach for molecules. The effective Hamiltonian approach may prove quite useful for molecular systems for which  electronic structure calculations are too expensive on existing quantum hardware.

\end{document}